\documentstyle[aps,preprint,multicol,epsf,epsfig]{revtex}

\newcommand{\beq}{\begin{equation}}

\newcommand{\eeq}{\end{equation}}
\newcommand{\barray}{\begin{eqnarray}}
\newcommand{\cc}[1]{c_{\vec{R}_#1,\sigma}}
\newcommand{\ccd}[1]{c^\dagger_{\vec{R}_#1,\sigma}}
\newcommand{\dd}[1]{c^\dagger_{\vec{R}_#1,\sigma_{\vec{R}_#1 }}}

\newcommand{\coned}[1]{c^\dagger_{#1,\sigma_{#1}}}
\newcommand{\bm}[1]{{\bf #1}}
\newcommand{\Tau}{{\cal{T}}}
\newcommand{\vd}{\vec{\delta}}
\begin{document}
\title{  Kinetic Antiferromagnetism  in the Triangular Lattice}
\author{Jan O Haerter and B Sriram Shastry }
\address{Physics Department, University of California,  Santa Cruz, Ca 95064 }
\date{\today}
\maketitle
\begin{abstract}
We show that the motion of  a single hole in the infinite $U$ Hubbard model with frustrated hopping
leads to  weak metallic antiferromagnetism of kinetic origin. An intimate relationship is demonstrated between  the simplest versions of this problem in 1 and 2 dimensions,  and  two of the most subtle many body problems, namely the Heisenberg Bethe ring in 1-d and the 2-dimensional triangular lattice Heisenberg antiferromagnet. 
\end{abstract}
\maketitle


The role of kinetic energy in  the theory of magnetism is crucial:   while virtual processes promote antiferromagnetism in insulators, as in the theory of superexchange, real ( or direct) kinetic processes usually promote ferromagnetism as a corollary of  metallicity,   as in the theory of double exchange\cite{herring,anderson1}. The Nagaoka Thouless (NT) theorem \cite{nagaoka,thouless} is of great  importance in providing a  rigorous mechanism for metallic ferromagnetism, notwithstanding the limitations of its context, that of  a single hole in the limit of infinite repulsion. In this work we  study the latter problem on certain  1-d and 2-d lattices  with {\em  electronic frustration}, a term defined by computing  the sign of the hopping amplitudes around the smallest closed loop of a lattice, in complete parallel to the more familiar spin counterpart. If the sign is negative then the lattice is said to be  electronically frustrated.   The NT theorem applies only to the non  frustrated cases, our focus is on the frustrated cases where not much is known reliably.   We find surprisingly   that in these cases the  {\em real  kinetic  processes  promote antiferromagnetism in the metallic phase}. 

Our immediate motivation for studying electronically frustrated lattices is to understand the physics of systems such as the recently found  sodium cobalt oxide system $Na_x Co O_2$\cite{levi}.  We present some analytical and numerical results for the one hole frustrated problem. Our main analytical tool is a novel reduction of the problem to an effective spin model, which yields  insights into  the physics. We have supplemented this with a numerical  
study using exact diagonalization of the effective spin problem, exploiting  the symmetries of the clusters. Following  the numerical work on the triangular lattice Heisenberg model (HM) by Bernu {\em et al}\cite{bernu} we compute the exact eigenvalues in different total spin sectors for clusters of various sizes, and this leads to ``towers of excitations''. These  help one to distinguish between spin liquid states and various ordered states, something that variational studies cannot quite do\cite{iordanskii}.    Our findings are consistent with a three sublattice broken spin symmetric ground state, very similar to that of the triangular lattice HM \cite{bernu,triangularhm}

We study periodic clusters of the infinite $U$ Hubbard model with $L$ sites and $N=L-1$  particles with 
 $H=- \sum_{i,j,\sigma} t_{i,j} \ccd{j} \cc{i}$, where $\cc{i}$ are Gutzwiller projected fermions with single occupancy constraint built into them.  Due to the periodicity, we may define a wave vector $\vec{k}$ for each wave function, and consider the action of $H$ in a fixed $\vec{k}$ subspace.  Let us locate the hole at site $\vec{R}_{i_0}$, and write a basis state $|\alpha\rangle= \dd{1} \dd{2}... \dd{N}|0\rangle$.  A generic state may be written as $\psi(\vec{k},\alpha) = \frac{1}{\sqrt{L}}  \sum_{\vec{r}} e^{(i \vec{k}.\vec{r})} T_{\vec{r}} | \alpha\rangle$, where $T_{\vec{r}}$ is a (spatial ) translation operator $ T_{\vec{r}} \ccd{i} T^\dagger_{\vec{r}}= \ccd{i+\vec{r}}$.
Exploiting the translation invariance of the Hamiltonian, the matrix element of $H$ in such a state is expressible in terms of an effective $\vec{k}$ dependent operator
\beq
\langle \beta |H_{eff}^{\vec{k}}| \alpha\rangle =\sum_{\vd} t_{\vd} \;  e^{- i \vd.\vec{k}} \langle \beta | T_{\vd}\; \cc{{i_0}-\vd} \ccd{i{_0}} |\alpha\rangle.
\label{heff} \eeq
 In this frame of reference, we hold the hole at a fixed site,  as the entire set of spins flows past it. While this reduction is suggestive, it is not yet equivalent to finding an effective {\em spin } exchange representation, since the 
translation operators carry the complexity of fermionic negative signs. We can make considerable progress in specific problems as follows. 

{\bf 1-dimension: }Let us first consider the  general 1-d  case with arbitrary range of hopping $t_r$.   Let  the hole reside at the $L$th site, so that the $L-1=N$  particles occupy sites $1\rightarrow N$ with some spin configuration $| \sigma_1,..\sigma_N \rangle= \coned{1} \coned{2}..\coned{N}|0\rangle$.  Operating with terms in $H$ shifting the hole to its left,  we find  $  t_{L,j} (-1)^{(L-j)} \coned{1}..\coned{j-1}\coned{j+1}..\coned{N}  c^\dagger_{L,\sigma_{j}}|0\rangle $. We next restore the hole to the $L$th site from the $j$th site, so we need to translate by $r=L-j$ units. This gives rise to negative signs that can be calculated readily, and the answer written down in terms of the permutation operator $P_{i,j} = \frac{1}{2} + 2 \vec{S}_i. \vec{S}_j$,  and  {\em spin translation operator} $\Tau= P_{1,2} P_{2,3} ...P_{N-1,N}$  acting as a cyclic permutation on a lattice of length $N$ as 
$$
H_{eff}^{k}=\sum_r (-1)^{r (L-r)} ( t_r e^{ - i k r} P_{r,r+1}...P_{2,3} P_{1,2} (\Tau)^r + hc).
$$
The sum is over the range of hopping,  thus the spin problem is defined on a  periodic ring of length $N$ rather than $L=N+1$.  In the simplest frustrated case we consider a model with nearest and second neighbor hopping $t_1 = t , \; t_2 = t'$, with both $t , t'>0$. This (railroad trestle) lattice may be considered a strip of the triangular lattice. For this case, and with $L$ odd, the effective  Hamiltonian is
\beq
H_k^{t,t'}=  t  e^{- i k} \Tau + t'  e^{ -2 i k} P_{1,2} (\Tau)^2 + hc , \label{h1d}
\eeq
the full spectrum is obtained by varying $k$. The hole has been eliminated, and we arrive at an  impurity bond model residing on the deleted $L-1$ site lattice.
The first term of Eq(\ref{h1d}) bodily translates all spin configurations,  hence all spin states remain degenerate.  The second term  discriminates between the configurations through the single exchange term $P_{1,2}$. 

\begin{center}
\begin{figure}
\epsfxsize=15cm
\epsffile{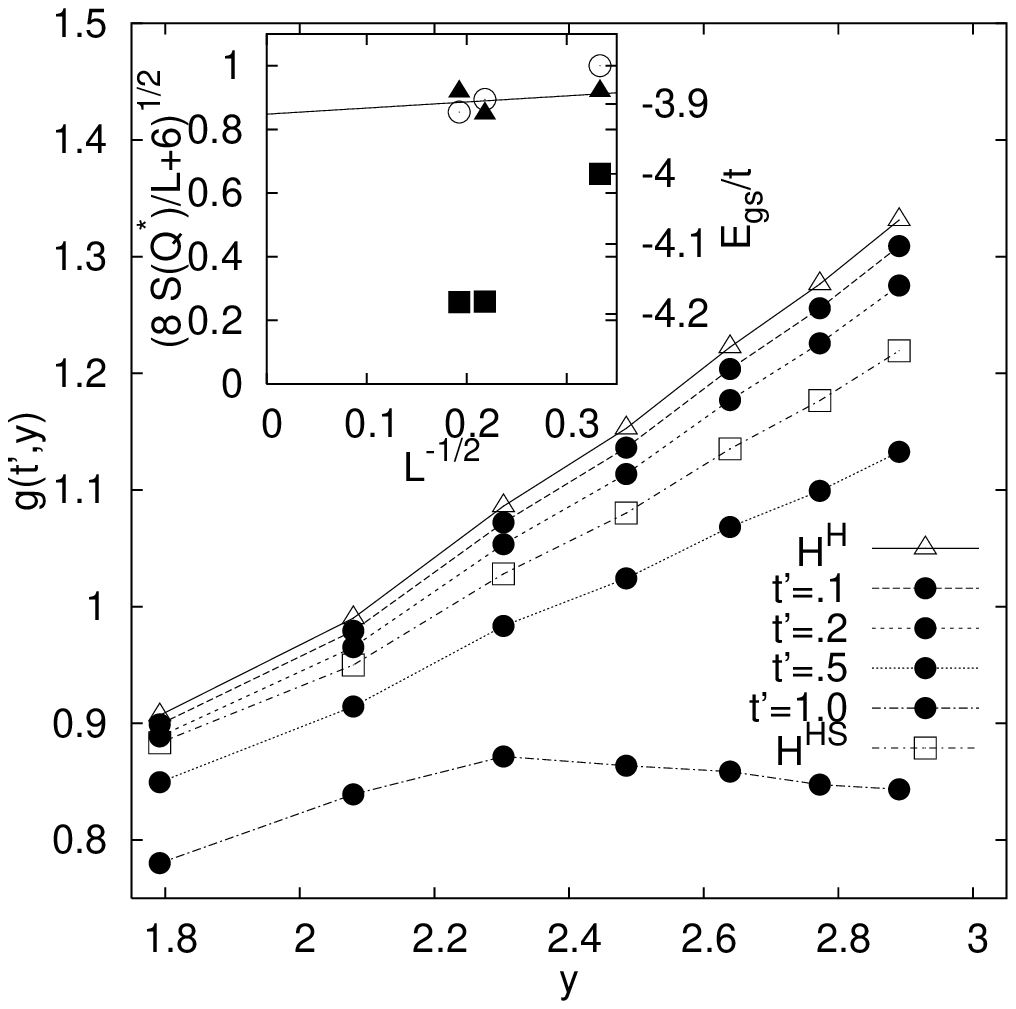}
\caption{ 1-d structure function  $g(y,t')$ of $H^{t,t'}$ for various $t'$ compared to Heisenberg ($H^H$) and Haldane-Shastry ($H^{HS}$) model. In the limit $t'\rightarrow 0$,  $H^{t,t'}$ correlations converge to those of $H^H$ .  
 {\bf Inset}:  2-d: Size dependence of the structure function on 2-d triangular lattice for Hubbard and Heisenberg \protect \cite{bernu} models denoted by solid triangles  and open circles. The solid squares are the ground state energy of $H_{eff}$ }
\label{gs}
\end{figure}
\end{center}

The ground state of Eq(\ref{h1d}) was  found numerically for rings up to $L=19$.  $L$ is taken to be odd to allow for the possibility of singlet ground states, and indeed in all cases, the ground state is a singlet\cite{tbcsinglet}.  The correlation functions are found by imposing pbc's with length $N$ \cite{notepbc}. These
 alternate in sign  and seem to decay  as a power law for small  $t'/t$, changing to a faster and possibly exponential decay at large $t'/t$. This  behavior is reminiscent of that of the HM with a second neighbor interaction, $H_h= J_1 \sum_n \vec{S}_n.\vec{S}_{n+1} + J_2 \sum_n \vec{S}_n.\vec{S}_{n+2} $.  Here it is known \cite{j1j2} that  the Bethe point $J_2=0$  has power law decay along with a logarithmic correction to the correlations so that asymptotically  $C(r)\equiv \frac{1}{N}\sum_{i=1,N} \langle \vec{S}_i\cdot \vec{S}_{i+r} \rangle= (-1)^r \{ \frac{A}{|r|}+ \frac{ B (\log{|r|})^\frac{1}{2}}{|r|} \}+O(1/r^2)$ \cite{notepbc}.  The logarithmic term arises from the umklapp term, and $B$ varies with $J_2$ within a  power law phase  persisting upto $J^*_2/J_1 \sim .24$ beyond which there is a gap in the spectrum. To compare with this behavior, we  compute  the structure function $g(y,t')=\sum_{r=1}^N (-1)^rC(r)$.   Plotted as a function of $y=\log{N}$ in the power law phase, it is expected to be a sum of two terms, one $\sim y$  from A, and another  $\sim y^\frac{3}{2}$ from  the logarithmic term B. For various values of $t'$ we show $g$ as function of $y$ in Fig(\ref{gs}). We also display the corresponding structure functions on the same lattice sizes for the nn HM ($J_2=0$) and the Haldane Shastry (HS) model\cite{haldaneshastry} which  has $B=0$. It is clear that the small $t'/t$ cases are very similar to the Heisenberg model, whereas the case of $t'/t$ large appear to be gapped.   We note that  in the limit $t'\rightarrow 0^+$, the structure function  $g$ approaches the corresponding value for the HM. Remarkably enough, the individual correlation functions $C(r)$ approach those for the nn HM to the available precision ( 6 decimal places)! Taking a closer look, we studied  the wave functions   for small clusters (upto L= 13).  We found that the ground state of the impurity model, when translated as $\sum_r \exp(i P r) (\Tau)^r |\mbox{\bf impurity gs}>$ with an appropriate (Marshall  sign) momentum P= 0  ($\pi$)  for N/2 even (odd),   becomes  {\em  the exact ground state of the Heisenberg model}. Based on these results  we conjecture that the $t'\rightarrow 0^+$ {\em impurity model ground state after translation becomes the  Bethe ground state of the Heisenberg  ring}. Finite $t'$ seems related to further ranged exchange on a smaller scale  than $t'$ ( i.e.  $o(t')$). We  thus find that the impurity spin model may be regarded loosely as the HM,  with a scaled down exchange $J^{eff}=t'/N$ to make meaningful comparisons of energetics. 

{\bf The Triangular Lattice}   In 2-dimensions the effective Hamiltonian  Eq(\ref{heff}) can again be reduced to a spin operator.  Let us imagine a two dimensional rhomboidal lattice with $N_1$ columns and $M_1$ rows ( $L= N_1 M_1$) and the topology of a torus. We consider a particular term where   the hole has hopped to the site $\vec{R}_{i_{0}} - \vd_i $, and the translation is along $\vd_i$ so as to restore the hole to the fixed site $\vec{R}_{i_{0}}$. It is enough to consider half the $\vd_i$'s corresponding to forward hops, the back hops contribute to the hermitean conjugate.   The translation is expressible as a specific permutation of the $L-1= (N_1 M_1 -1 )$ variables $\vec{R}_{l}$ written in some specific order. This permutation can be decomposed into 
$c_i$ cycles, each  of length $l_m$ so that $\sum_m l_{m}= L-1$. The translation operator $T_{\vd_i}$ accumulates phase factors of $(-1)^{l_m-1}$ from each cycle relative to the pure spin translation operator $\Tau_{\vd_i}$,
 and one extra minus sign arises from the row containing the hole,  so that the overall phase factor is $-(-1)^{\sum_m (l_{m} -1)} = (-1)^{ L  - c_i}$ , and hence 
\beq
H^{k}_{eff}= \sum_i \{ t_{\vd_i} (-1)^{L - c_i} e^{- i \vec{k}.\vd_i} \Tau_{\vd_i} + h.c. \}.
\eeq
For the triangular lattice  we note that $\vd= {\hat{x}}$ gives $c_i=M_1$  i.e. the number of rows, $\vd=\frac{1}{2} {\hat{x}} 
+\frac{\sqrt{3}}{2} {\hat{y}}$ gives $c_i= N_1$  i.e. the number of columns. In the case of the third hop $\vd=- \frac{1}{2} {\hat{x}}  +\frac{\sqrt{3}}{2} {\hat{y}}$, $c_i$   depends upon the relative prime-ness of $N_1$ and $M_1$ but is trivial to compute for any  given cluster once and for all.  The translations $\Tau$ act only on the spin labels, and satisfy $\Tau_{\vd_i} c^\dagger_{\vec{R},\sigma_{\vec{R}}}  \Tau^\dagger_{\vd_i} = c^\dagger_{\vec{R},\sigma_{\vec{R}+\vd_i}}$ for $\vec{R} \neq \vec{R_{0}} + \vd_i$, and $\Tau_{\vd_i} c^\dagger_{\vec{\vec{R_{0}} + \vd_i},\sigma_{\vec{R_{0}} + \vd_i}}  \Tau^\dagger_{\vd_i} = c^\dagger_{\vec{\vec{R_{0}} + \vd_i},\sigma_{\vec{R}-\vd_i}}$, i.e. is a unit spin translator along the $\vd_i$ direction for all sites except the one nearest to the hole where it shifts by two units.

To gain insight into the type of magnetic ordering in the triangular lattice we have performed exact diagonalization of small clusters \cite{clustersfig}.
We exploit the translation and rotation invariance by working with the deleted lattice in the subspace with $S_z=0$. We also use time reversal invariance so that
under  a global transformation $\sigma_j \rightarrow -\sigma_j$,  its eigenvalues are $\kappa=\pm 1$. We find   that if $(L-1)/2$ is odd (even)  the even spin states   have $\kappa= - 1 (1)$, the odd spin states  the opposite value. 

\begin{center}
\begin{figure}
\epsfxsize=15cm
\epsffile{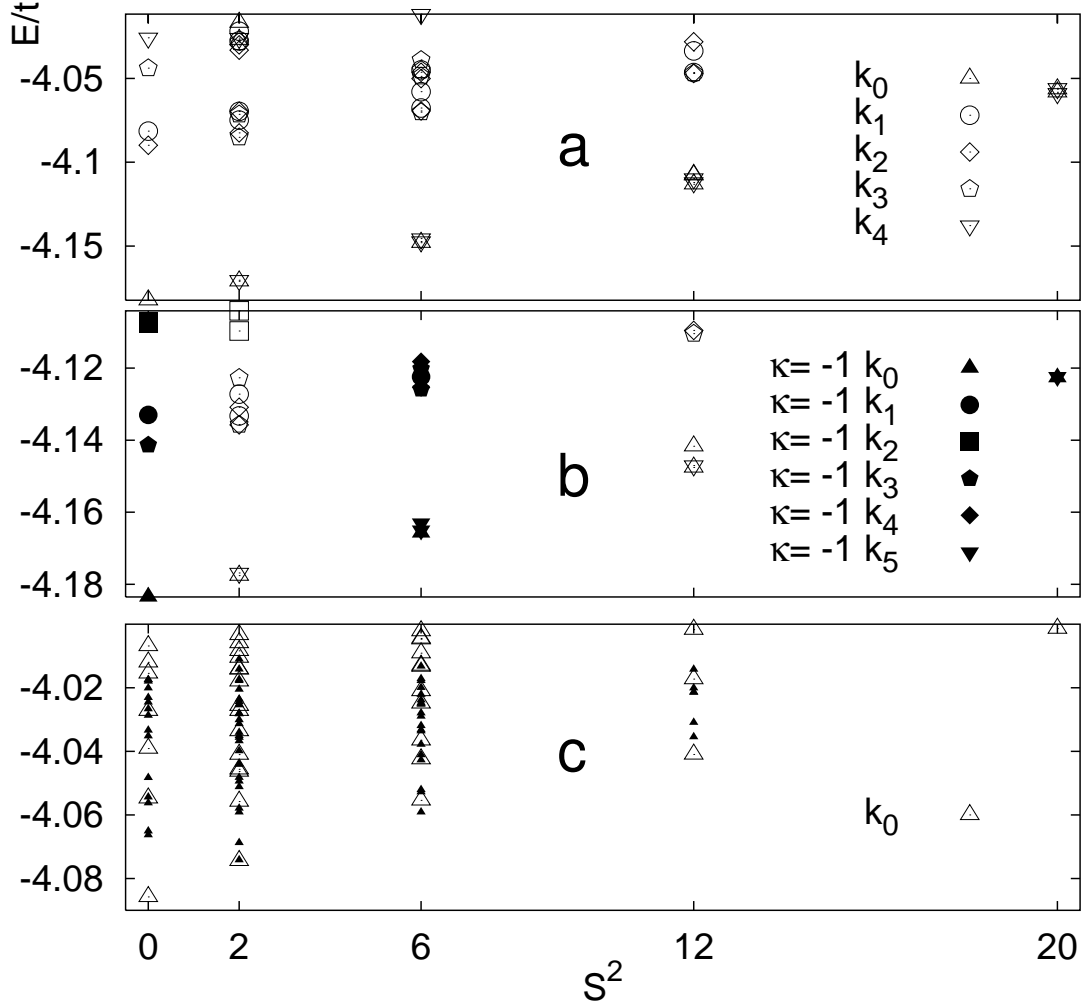}
\caption{a (b):Clusters with 21 (27) sites. Symbols correspond to different subspaces: $k$ - momentum eigenvalue numbered in increasing magnitude of $k$, b: solid (open) symbols correspond to $\kappa=-1$ ($\kappa=1$) c: The frustrated 21 site cluster, zero momentum states ($k_0=0$) emphasized}
\label{towers2127}
\end{figure}
\end{center}

 We  contrast systems either supporting or frustrating 3-sublattice order. In particular, we choose systems of $9$, $21$ and $27$ sites which support  3-sublattice order\cite{clustersfig}.   The $3\times 3$-cluster displays high spatial symmetry which is reflected by a $7$-fold degenerate ground state, one of which is a singlet which lies in the ($\bm{k}=0$, $\kappa=1$)-sector. Due to its small size and geometry, in this cluster, every nearest neighbor (nn) is also a second neighbor and the third neighbor is the site itself. Thus, 3-sublattice order is forced by the boundary conditions in this cluster. We found a vanishing of the fluctuation of  the operator $Q=\sum_{<{i},{j} >}(\vec{S}_{{i}} \cdot \vec{S}_{{j}} - n_{\bm{i}} n_{\bm{j}}/4)$  in the singlet ground state. This curious result implies that within the tJ-model, both the contributions to the Hamiltonian share a common ground state for this cluster. In terms of the deleted lattice, the ground state of $H_{eff}$ is exactly the ground state of the nn HM. Next, we compute the spectra of the $21$ and $27$ site clusters.  We found that  the states corresponding to $\vec{k}=\vec{0} \mbox{ or } \vec{Q}^*\equiv \frac{4 \pi }{3} \hat{x} $ were found to be lower in energy and nearly degenerate.  This can be attributed to a unit cell tripling, as one expects in case of 3-sublattice order. We find in Fig(\ref{towers2127}) a systematic behavior of the excitations, in close parallel to those for the triangular lattice HM\cite{bernu}.
One can define a ``moment of inertia''\cite{bernu} $I$ as the inverse of the  slope of the line joining the bottoms of the different $S_{tot}$ towers of excitations. Our moment of inertia $I\sim L^2$ as compared to the Heisenberg value $\sim L$, again understandably in view of the extensivity of the latter model, and suggests $J_{eff} \sim t /L$.
A noteworthy feature is the striking ``subgap'' in each tower, this separates the ground state from the excitations that  start out sparsely and then seem to form a continuum. A very similar feature in the HM  has been identified with the magnons\cite{bernu} with an energy scale $\omega = c |k| \sim 1/\sqrt{L}$, and seems equally relevant here.
In the $21$-site cluster the states corresponding to  three vectors $k_1$, $k_2$ and $k_3$ are all very close in energy. This can be understood in terms of a $1^{st}$ BZ diminished in size by $1/3$ as these three momentum vectors are placed in equal distances from the corners of this new $1^{st}$ BZ. In the $27$-site cluster, the picture is very similar. Again, the $\vec{k}=\vec{0} \mbox{ or } \vec{Q}^* $ states are clearly separated from the rest of the spectrum. The family of first excited states is constituted by the momenta $k_1$, $k_3$ and $k_4$, which are now vectors that lie exactly on the corners of the diminished $1^{st}$ BZ. In this cluster, a second family of excited states emerges, which is populated by states belonging to the $k_2$ subspace. This vector lies in the center of the neighboring diminished $1^{st}$ BZ in y-direction of $k_0$. This point corresponds to a state of half the wave-number of the classical N\'eel state and is the least 'compatible' with the 3-sublattice scenario, thus leading to the highest excitation energy.

We also studied system of $15$ and $21$ sites, which frustrate the three sublattice order through their particular choice of boundary conditions\cite{clustersfig}. We  found that the energies of the frustrated clusters  are considerably higher than for the unfrustrated ones. For 21 sites we can compare these directly: they are respectively  $-4.08577 \mbox{ and } -4.18233$, thereby showing the clear preference for three sublattice order. In plot (c) of  Figure(\ref{towers2127}) we  compare the tower of states obtained for this frustrated  21 site cluster. The regular features described above are now perturbed, the zero-momentum states no longer constitute the ground states of the different spin sectors and the quasi-degeneracy can no longer be identified. This scenario is similar to that obtained by a HM with an additional 4-site ring exchange term \cite{misguich_multiplespin}.

 We  also calculated the  structure function $S(\vec{q})=\sum_{i,j}\exp{(i\vec{q}\cdot (\vec{r}_{\bm{i}}-\vec{r}_{\bm{j}}))}<\vec{S}_{i} \cdot \vec{S}_{j}>/L$.  It is useful to identify $v=\sqrt{S(\vec{Q}^*)/(L+6)}$ as the physical order parameter, with the same normalizations as in the HM\cite{bernu}. The clusters compatible with 3-sublattice order contain the  wave vector $\vec{Q}^*$-value, unlike  the frustrated clusters. While the compatible clusters show a clear peak for $S(\vec{Q}^*)$, the frustrated clusters still show a peak at the $\bm{q}$-vector closest to $\vec{Q}^*$. Thus, 3-sublattice order is likely to be dominant in arbitrary systems and is not an artifact of a biased choice of boundary conditions. For the compatible systems we plot in the inset of  Fig. (\ref{gs}) $v$ vs. $L^{-1/2}$. and  compare with results for the HM obtained in \cite{bernu},   normalizing to the maximum possible value of $S(\vec{Q}^*)$. Extrapolation  to infinite system size suggests a substantial finite intercept ( $\sim0.849$). This shows  strong similarities with the studies on the HM\cite{bernu}, and   suggests N\'eel long range order originating solely from the motion of a single hole in a spin background.

In Fig (\ref{gs} inset) we also plot the ground state energy. The small system size and high symmetry of the 9 site cluster leads to a fairly large ground state energy while the 21 and 27 site clusters appear to show rapid convergence, we estimate  $E_{gs}=-4.183 \pm 0.005$. On the square lattice, this swift convergence has also been observed by Poilblanc et al. \cite{poilblanc}.  Finally, the impurity model allows us to study the nature of the {\em spin texture} surrounding the hole\cite{clustersfig}.  In the (unfrustrated) hexagon surrounding the hole, we find substantial alternating  antiferromagnetic  order $\langle \vec{S}_i \cdot \vec{S}_j \rangle\approx -0.34$  similar in magnitude and nature to that observed for a HM on a square lattice \cite{young}. This local impurity-order rapidly transitions into three-sublattice antiferromagnetic correlations at greater distances from the hole. Thus, the hole can be seen as a moving impurity around which spins tend to line up antiferromagnetically.

In conclusion, we have argued for  {\em kinetic  antiferromagnetism} on frustrated lattices. The preference for antiferromagnetism arises from  the subtle phase dependence of the kinetic motion. Hole hopping  mimics the effect of  antiferromagnetic exchange energy, with  the scale  of a single impurity exchange bond  per hole.
The remarkable mapping to the Bethe ring in 1-d and close similarity with the HM  on  triangular lattice highlight  this subtle phenomenon. Our effect should be most prominent in situations where superexchange is negligible.   
At least qualitatively used, as in  Ref\cite{anderson2}, our findings can thus be interpreted as leading to {\em weak antiferromagnetism} with $J_{eff}\sim + c x |t|$ in the case of the frustrated   lattices, where $x$ is the hole concentration.

\begin{acknowledgments}
This work was supported by NSF-DMR 0408247. We are grateful to O. C\'epas,  Z. Fisk  and  A. P. Young for  helpful comments.

\end{acknowledgments}

\newpage
\newpage
 
\begin{center}
{\bf Supplementary Plots}\\
{\bf 3 sublattice-order preserving clusters}\\
Distinct colors correspond to different sublattices.
\end{center}

\begin{figure}
\epsfxsize=13cm
\epsffile{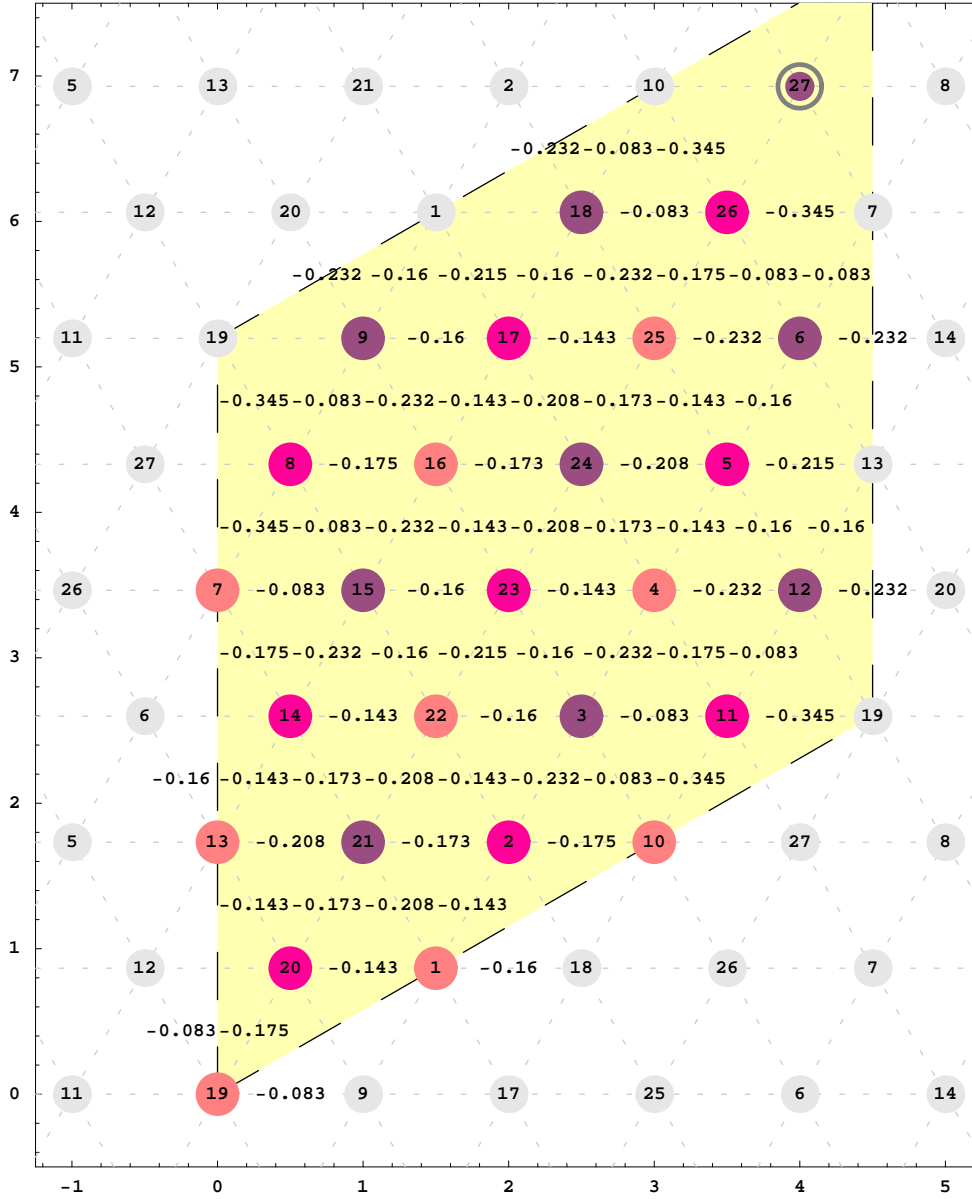}
\caption{The largest (27 site) cluster used in our exact diagonalizations: localized hole on site 27, decimal numbers between site $i$ and $j$ show correlation function $<S_i\cdot S_j>$, note behavior of correlation function on hexagon surrounding the hole }
\label{gs27sites}
\end{figure}
\newpage
\begin{figure}
\epsfxsize=15cm
\epsffile{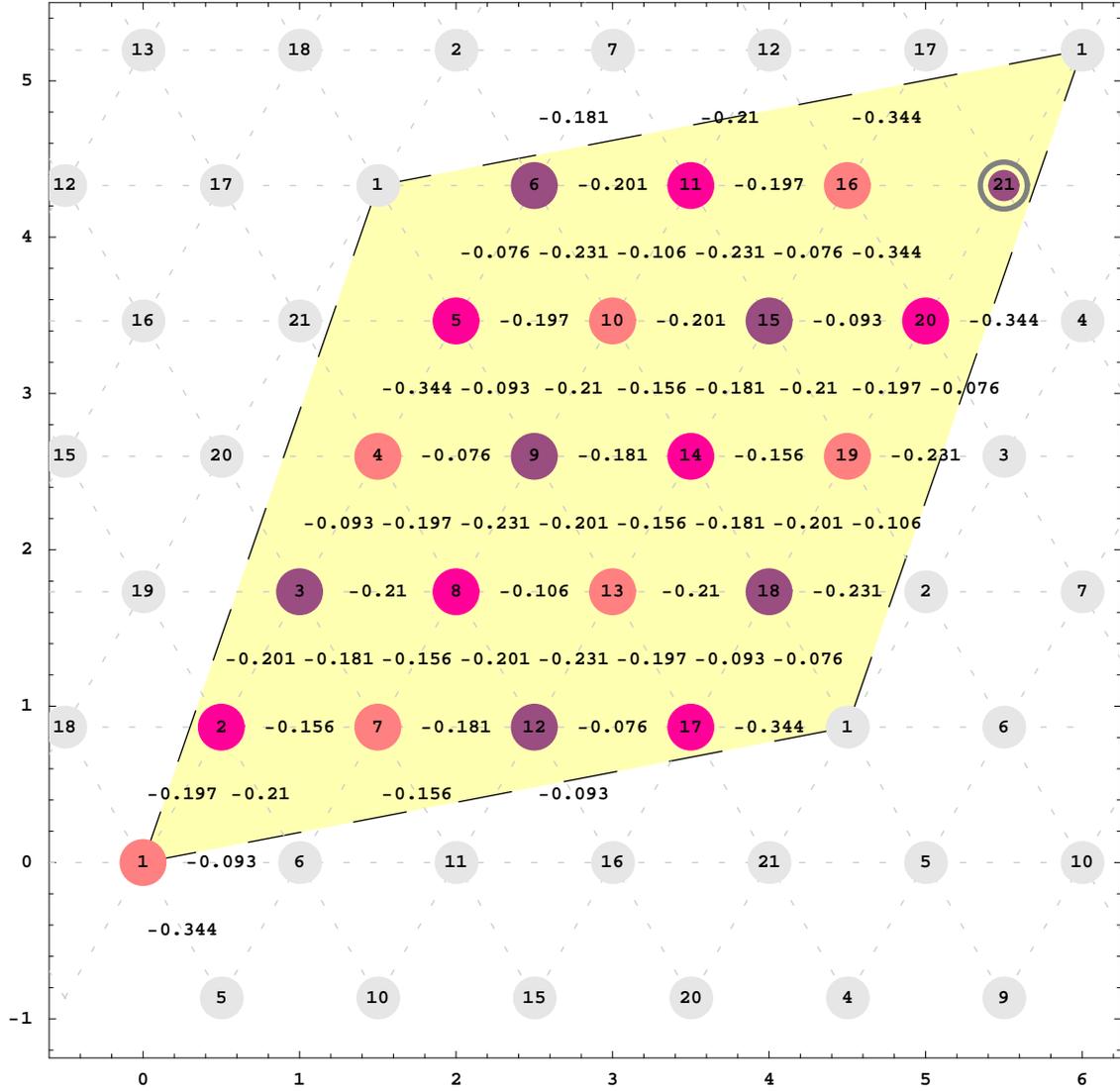}
\caption{The 21 site cluster. Note that this cluster maps onto a 1D ring with hoppings $t_1$, $t_4$ and $t_5$ by moving along the directions of numbered sites and using periodic boundary conditions }
\label{gs21site}
\end{figure}
\newpage
\begin{figure}
\epsfxsize=15cm
\epsffile{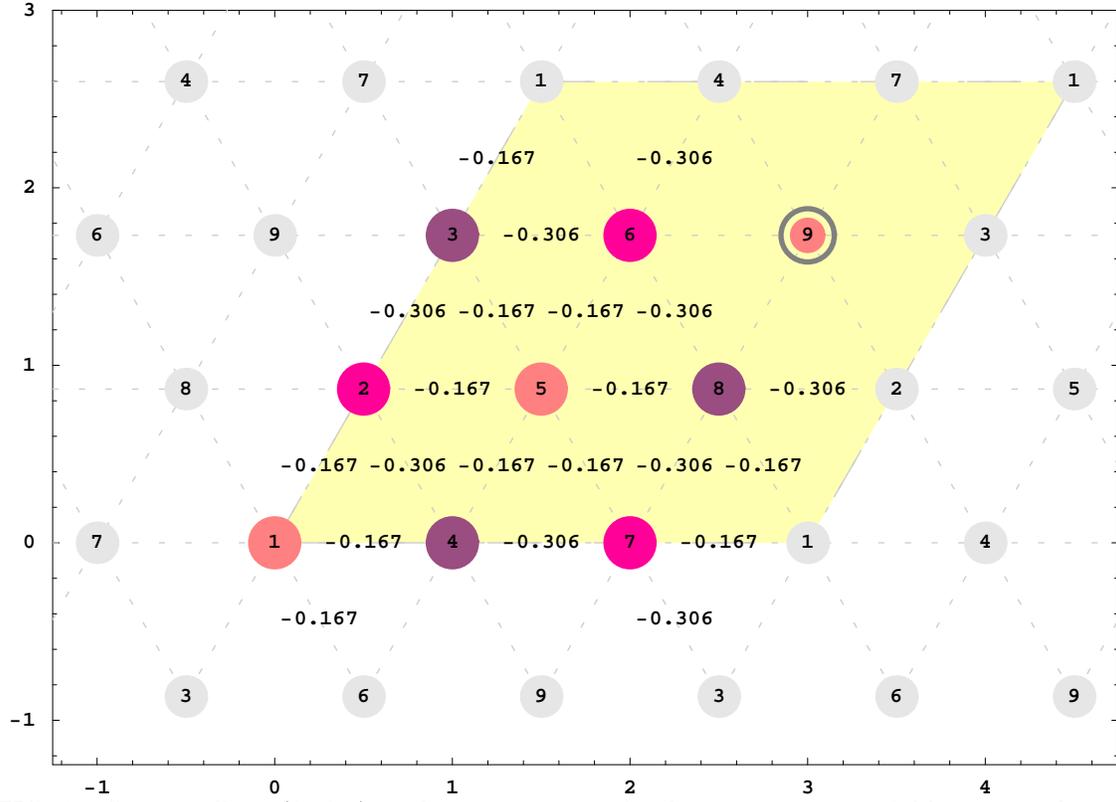}
\caption{Our smallest (9 site) and most symmetric cluster, nearest neighbors are also second neighbors and third neighbors are the same as the site itself. The high symmetry leads to a 7-fold degenerate ground state. }
\label{gsninesite}
\end{figure}

\newpage
\begin{center}
{\bf 3 sublattice-order frustrating clusters}
\end{center}
\begin{figure}
\epsfxsize=15cm
\epsffile{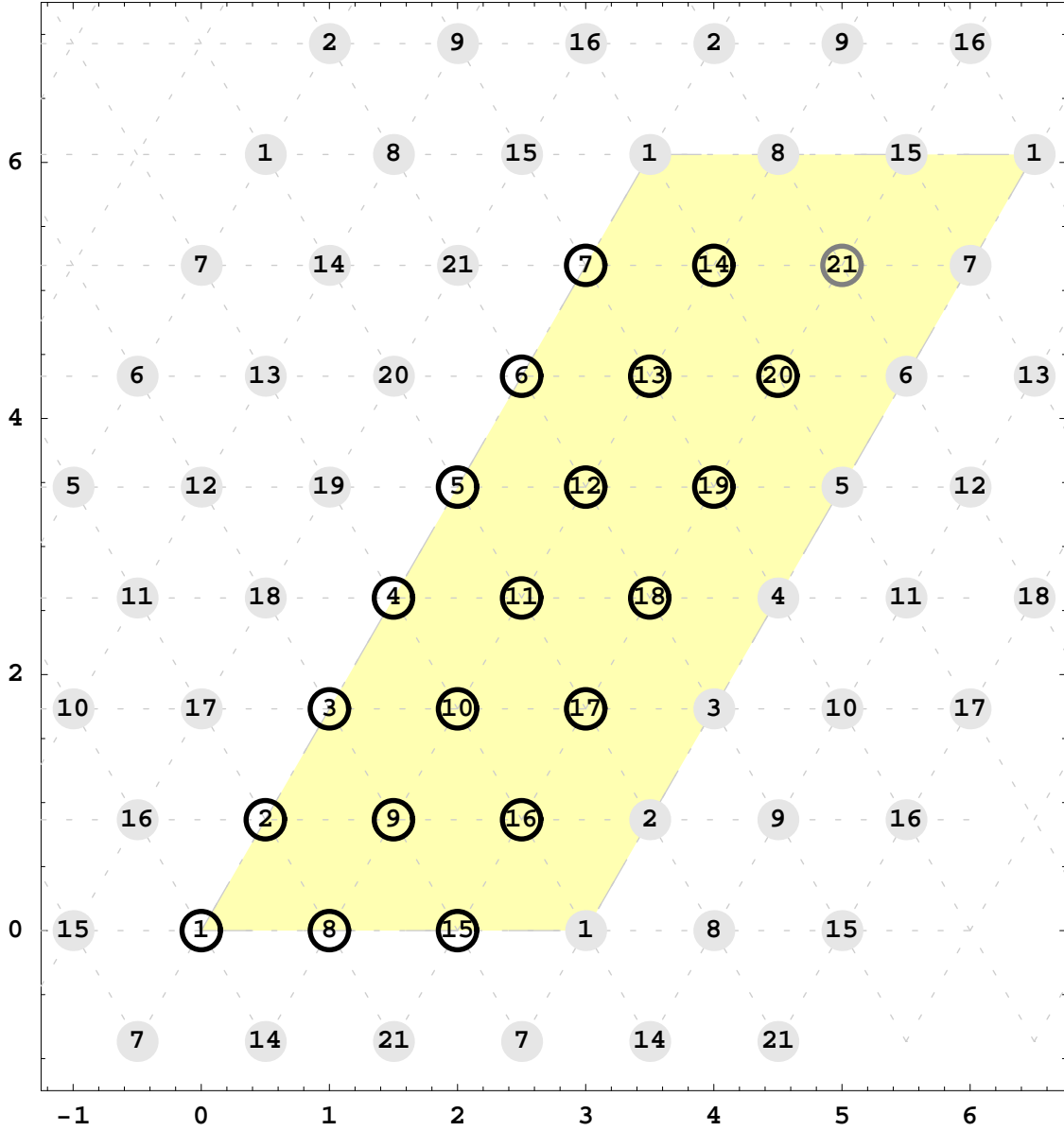}
\caption{The 21 site cluster frustrating 3-sublattice order. Check this by moving along the direction 1,2,3,... and placing 3 alternating classical N\'eel spins on consecutive sites, when arriving at the boundary a discontinuity arises.}
\label{gsfrustrated}
\end{figure}

\newpage

\begin{figure}
\epsfxsize=15cm
\epsffile{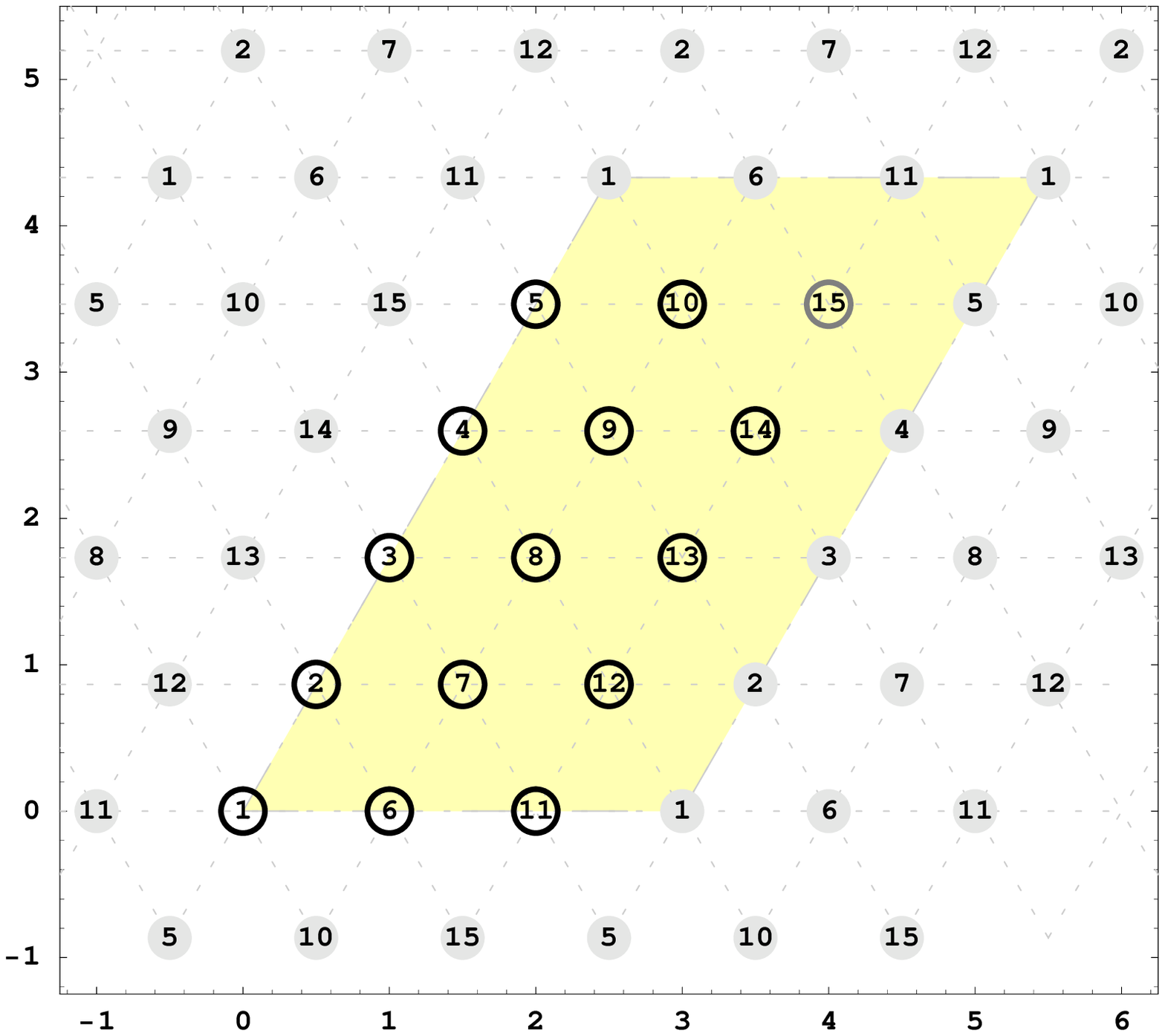}
\caption{The 15 site frustrated cluster }
\label{gs15site}
\end{figure}

\begin{references}
\bibitem{tbcsinglet} The ground state occurs at $k=0$ for L=3,7,11.. and at $k=\pi$ for L=5,9,13... Strictly speaking $k=\pi$ is not allowed in these cases,  we therefore insert an appropriate flux in the  ring to shift the momenta.
\bibitem{notepbc} The Hubbard correlations actually peak at a  wave vector different from $\pi$ by terms of $O(1/L)$ due to the slight doping. The correlations functions found by imposing PBC's of length $N$ are more systematic, peak exactly at $\pi$ and   differ from those of  the Hubbard case by known terms of $O(1/L)$.
\end{references}
\end{document}